# Direct Amplification of Sound by Light

**Chunnong Zhao, Li Ju, Haixing Miao, Slawomir Gras, Yaohui Fan &
David G. Blair**

School of Physics, The University of Western Australia, Crawley,
Western Australia, 6009 Australia

A new device, the opto-acoustic parametric amplifier (OAPA), is introduced, which is analogous to an optical parametric amplifier (OPA). The radiation pressure reaction of light on the reflective surface of an acoustic resonator provides nonlinearity similar to the Kerr effect in the OPA. The OAPA can be tuned to operate in two regimes. In the positive gain regime, an acoustic signal is directly amplified. In the negative gain regime, we show that the acoustic resonator can be cooled to the quantum ground state.

Macroscopic devices based on three frequency interactions of electromagnetic waves have provided powerful tools across a range of disciplines from radio astronomy to quantum optics. In the 1960's the non-linear properties of varactor diodes were used to create extremely low noise non-degenerate parametric amplifiers for microwaves [1]. More recently non-linear optical crystals have been used to create optical parametric amplifiers and oscillators (OPAs and OPOs) with an enormous range of applications. Here we introduce the opto-acoustic parametric amplifier (OAPA). This device uses the intrinsic non-linearity provided by the radiation pressure reaction of light on a reflective acoustic resonator to create a system closely analogous to the OPA except that one channel is acoustic. The OAPA utilises a resonant interaction between three high quality factor resonators, (two optical, and one acoustic) which give rise to a strong opto-acoustic coupling, direct positive or negative acoustic gain and acoustic signal transduction to an optical frequency. We demonstrated an OAPA using an 80m optical cavity [2]. Such systems can be reduced to 10 cm scale if a coupled optical cavity is introduced to enable tuning of the Guoy phase. We show here that in a high negative gain mode an OAPA can cool relatively high mass acoustic resonators to their quantum ground state using only low levels of optical power. The third resonant mode avoids optical noise that limits cooling using single mode optical cavities. In close analogy with the OPA, the OAPA can create entangled pairs of phonons and photons [3], which has many potential applications in quantum measurement and quantum computing.

Three-mode radiation pressure mediated opto-acoustic parametric interactions were predicted by Braginsky *et al* [4,5] in the context of long optical cavities for gravitational wave detectors. Zhao *et al* [6] modelled interactions taking into account the 3D acoustic mode structures of cavity mirrors and the optical cavity mode shapes, showing that the phenomenon was likely to be observable with current technology.

Three mode parametric interactions can be considered as a scattering process between a high intensity carrier $\omega_0$ and a pair of acoustic and optical modes described by the diagrams in Figure 1 [7]. The frequencies $\omega_0$ and $\omega_1$ are eigenmodes of an optical cavity for which one mirror is the surface of an acoustic resonator of frequency $\omega_m$. In (a) a photon of frequency $\omega_0$ is scattered, creating a lower frequency photon of frequency $\omega_1$ and a phonon of frequency $\omega_m$, causing the occupation number of the acoustic mode to increase. In (b) a photon of frequency $\omega_0$ is scattered by a phonon to create a higher frequency photon, thereby reducing the acoustic mode occupation number. For the configurations considered in this letter both processes are separately accessible. The time reversed processes must be considered when the occupation number of $\omega_m$ is small.

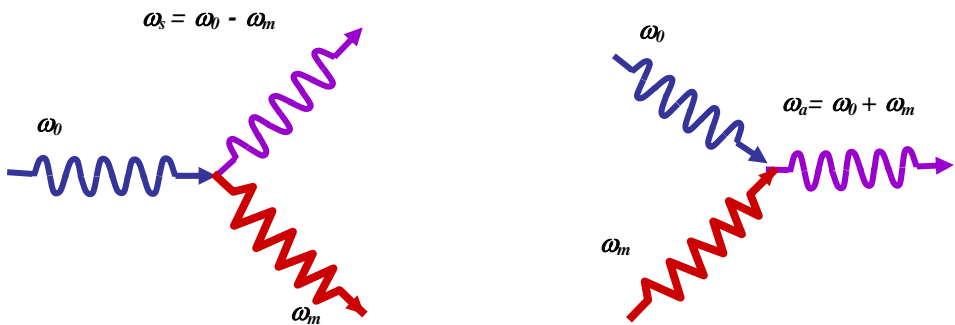

Figure 1  Schematic diagram of the parametric phonon – photon interactions in an OAPA. In (a) a photon of frequency $\omega_0$ is scattered into a lower frequency photon $\omega_1$, and a phonon of frequency $\omega_m$, while in (b) the scattering creates a higher frequency photon $\omega_1$, which requires the occupation number of the acoustic mode to reduce. Arrows indicate the direction of power flow when the occupation number of the $\omega_0$ mode is high.

For efficient 3-mode parametric interaction two conditions must be met simultaneously. Firstly, the optical cavity must support eigenmodes that have a frequency difference approximately equal to the acoustic frequency. Secondly, the optical and acoustic modes must have a suitable spatial overlap.

The effect of parametric scattering can be characterised by a dimensionless parametric gain, $R$. The gain depends on the product of three quality factors: those of the optical modes $Q_0$ and $Q_1$ and of the acoustic mode $Q_m$. For a single interacting set of modes, the gain $R$ is given by [5]

$$R = \pm \frac{8 P_{in} Q_0 Q_1 Q_m}{L^2 \omega_0 \omega_m^2} \frac{\gamma B_1 / m_{eff}}{1 + (\Delta\omega/\delta_1)^2} \quad (1)$$

Here $P_{in}$ is the incident laser power; $\omega_0$ is the frequency of the TEM$_{00}$ mode of the cavity; $m_{eff}$ is the mode effective mass of the acoustic resonator; $L$ is the length of the cavity; $\Delta\omega = \Delta - \omega_m$, where $\Delta = |\omega_0 - \omega_1|$; and $\delta_1 = \omega_1/2Q_1$ is the half linewidth (or damping rate) of the high order optical mode. The factor $B_1$ measures the spatial overlap between the electromagnetic field pattern and the acoustic displacement pattern defined in reference [5]. The coupling factor $\gamma = T/loss < 1$, where *loss* is the total round-trip loss of the cavity and $T$ is transmission of the input mirror. If the cavity loss is dominated by input mirror transmission, then $\gamma \sim 1$.

Because optical and acoustic quality factors can be very high, the parametric gain $R$ can be very large. For positive $R < 1$ the amplitude of the acoustic mode is increased by a factor *1/(1-R)* [8]. If $R > 1$, the OAPA becomes an oscillator, for which the acoustic mode amplitude increases exponentially with time until non-linear losses lead to saturation. For negative values of $R$, the OAPA extracts energy from the acoustic mode and reduces its effective temperature. The mode temperature, Q-factor and damping rate are changed form their original value $T_0$, $Q_m$ and $\delta_m = \omega_m/2Q_m$ to $T_{eff}$, $Q_{eff}$ and $\delta_{eff}$, given by $T_{eff} = T_0 \delta_m / \delta_{eff}$, $Q_{eff} = \omega_m / 2\delta_{eff}$ and $\delta_{eff} = (\delta_1 + \delta_m)/2 - real(\sqrt{(\delta_1 - \delta_m)^2 + 4R\delta_1\delta_m})/2$ [8]. The cooling limit is set by $T_0 \delta_m / (\delta_m + \delta_1)$ when the parametric gain reaches $R_{max} = -(\delta_1 - \delta_m)^2 / 4\delta_1 \delta_m$. This limit arises because the acoustic mode energy dissipation rate to the optical mode is limited by the optical mode energy dissipation rate to the environment. Otherwise the high order mode couples back into the acoustic mode. In equilibrium, the effective rates of energy dissipation for the optical and the acoustic modes are the same; energy dissipation is shared equally between the two modes. For $R < |R_{max}|$ the effective temperature of the acoustic modes is $T_{eff} \sim T/(1-R)$.

The cooling of macroscopic resonators to the ground state using detuned 2-mode interactions has been proposed and numerous experiments have demonstrated cooling [9,10,11,12]. However Marquardt *et al* [13] have shown that while resonators can be cooled for carrier power tuned to the lower 3dB point of the optical cavity resonance, the cooling of resonators to arbitrarily small phonon numbers can only occur in a regime called resolved sideband cooling, where the detuning is much larger than the cavity linewidth. In this regime the power build up is low, and relatively high incident power is required. This situation exacerbates the effects of laser intensity noise and phase noise which already limit cooling experiments. Laser noise drives the acoustic resonator with a fluctuating force that increases the resonator effective temperature. The effective temperature due to the laser technical noise heating is proportional to the square of the cavity input power [14]. Comparing the OAPA cooling to resolved sideband cooling, the OAPA is doubly resonant, and the input power is greatly reduced for the same circulating power. For example, with a modest cavity build-up of $10^2$ the laser noise requirement for the OAPA is $10^4$ times less than that for resolved sideband cooling.

To confirm the physics of the OAPA, we used an 80m suspended optical cavity with about 4W of 1064nm incident power and cylindrical sapphire mirrors of mass ~ 5.5kg [2]. For this experiment we tuned the cavity using an electrically heated intra-cavity fused silica thermal tuning plate [15]. This plate acts as a variable focal length lens allowing the cavity g-factor to be tuned from 0.87 to 0.99. For $g$=0.934, the frequency difference $\Delta$ between TEM$_{00}$ and TEM$_{01}$ modes become resonant with acoustic modes at 160kHz, leading to Lorentzian peaks in the detected TEM$_{01}$ mode power. We also confirmed linear transduction and correlation between the optical and acoustic signals by comparing direct readout of acoustic modes excited in $M_2$ with the $TEM_{01}$ signal [2].

From equation (1) the maximum achievable parametric gain is $R=8P_{in}Q_0Q_1Q_m/L^2\omega_0\omega_m^2 m_{eff}$. The positive acoustic amplification regime where $R <\sim 1$ is easily achieved. For example, $R=1$ is achieved for a cavity finesse of 5,000 ($Q_0 \approx Q_1 \sim 10^{10}$), and $Q_m=10^5$, using 1mW of a YAG laser incident power, $L$=1m, $m_{eff}$=1mg and $\omega_m=2\pi\times10^6$. In this case the OAPA acts as direct acoustic amplifier with gain $1/(1-R)$. An acoustic signal is simultaneously transduced to frequency $\omega_1$ with power gain $\omega_1/\omega_m \sim 10^9$. In the negative gain regime the OAPA acts as an acousto-optic transducer with strong cold damping. To achieve quantum ground state cooling both optical and acoustic Q-factors must be high. Kleckner et al [16] have demonstrated optical cavities with small mirrors and finesse $\sim 5\times10^4$.

Because optical coatings have high acoustic loss, it is necessary that the coating mass contribute a small fraction of the resonator mass, and that the part of the resonator supporting the mirror coatings be physically separated from the part experiencing large elastic deformation. The multilayer $SiO_2/Ta_2O_5$ coatings required to achieve low optical loss have relatively large and roughly temperature independent mechanical loss $\sim 10^{-4}$ [17]. For micro-mirrors in the µg range as used in 2-mode cooling experiments [9], the resonator mass is comparable to the mass of the lossy coating, and the Q-factor is limited by the Q-factor of the lossy coating materials. Three-mode cooling makes it possible to cool much larger mirrors (mm scale) so that coating mass fraction is small. Generally the Q-factor is determined by [17] $Q^{-1}=Q_{int}^{-1}+Q_{coating}^{-1}\times(\Delta E/E)$, where $\Delta E$ is the strain energy stored in the coating and E is the total strain energy of resonator mode concerned.

We now show that a small scale OAPA can cool a 1mm scale mini-resonator (1mg, 1MHz) to the quantum ground state. This requires a parametric gain $R=-kT_0/\hbar\omega_m$. For $T_0 \sim$4K, this corresponds to $R\sim 10^5$. We find that it is not possible to simultaneously tune the mode shape, mode size and mode frequency of the mini resonator cavity without an additional mirror and lens as shown in Figure **2b.** The mini-acoustic-resonator serves as the end mirror $M_2$. Mirrors $M_0$ and $M_1$ with lens $L_1$ creates an effective mirror, which forms a coupled cavity to tune the Guoy phase of the $TEM_{01}$ mode [18] to the required phase difference of $\Delta\phi=4\pi L\Delta/c$, so that $TEM_{00}$ mode and $TEM_{01}$ mode are both resonant in the cavity with the end mirror $M_2$. With careful selection of the mirror parameters we can achieve beam spot radius of 0.12mm on the resonator mirror, $TEM_{00}$ mode power build up of 237, and appropriate Guoy phases. Because of the frequency dependent reflectivity of the effective mirror the optical Q-factors for $Q_0$ and $Q_1$ are different: $8.8\times10^9$ and $1.2\times10^{10}$ respectively. We find that frequency $\omega_1$ can be tuned from 1 MHz above $\omega_0$ (corresponding to a negative R) to 1 MHz below $\omega_0$ (corresponding to a positive *R*) by moving the effective mirror $M_2$ about a few mm. To be tuned to within a cavity linewidth of a given acoustic frequency requires the mirror position to be adjusted within a few μm.

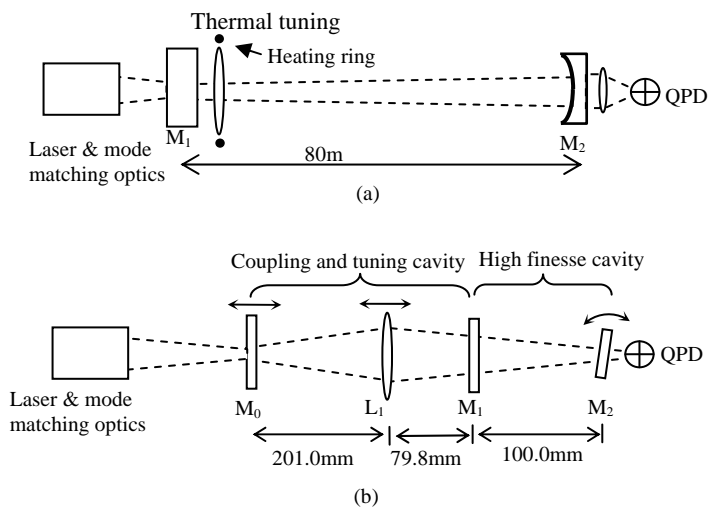

Figure 2 Long and short cavity implementations of the OAPA. (a) The long cavity implementation has been demonstrated [2]. The thermal tuning plate varies the cavity g-factor to tune the $TEM_{01}$ frequency difference to an acoustic mode of mirror $M_2$. (b) The short cavity implementation of the OAPA with a coupled cavity. In both (a) and (b) cases a differential quadrant photodetector (QPD) is configured to detect the beat signal between the $TEM_{00}$ and $TEM_{01}$ modes. The differential readout selects the $TEM_{01}$ mode and rejects common mode noise in the $TEM_{00}$ mode.

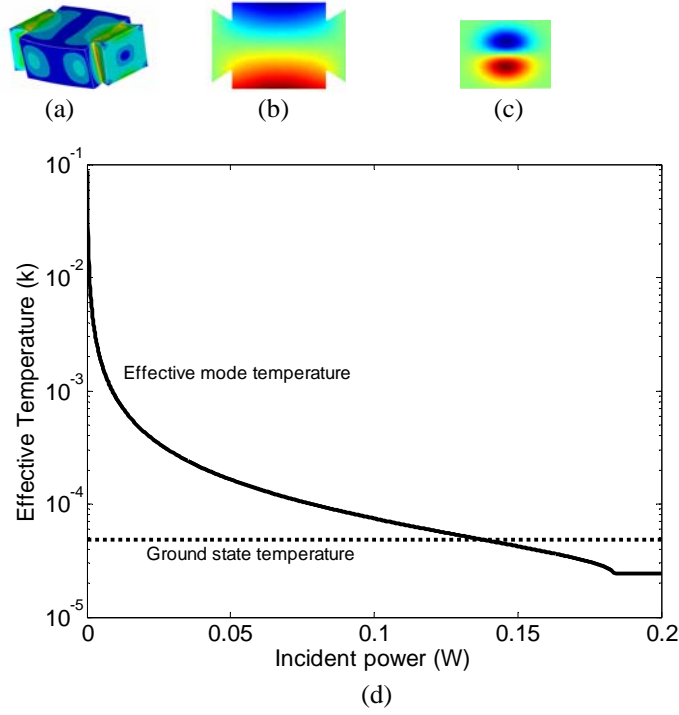

(a)  (b)  (c)

(d)

Figure 3 Mini-resonator design and the effective temperature achievable. (a) A silicon mini- resonator torsional mode with resonant frequency ~1 MHz. Dimensions ~ 1 mm × 0.8 mm × 0.5 mm, and mass ~1 mg. The contours show the strain amplitude, which is concentrated in the spindle ends and minimum at the centre where the $\omega_l$ mode interacts. (b) Displacement distribution of the surface of the mini-resonator torsional mode, showing a uniform displacement gradient along a central vertical axis. (c) The optical $TEM_{01}$ mode amplitude distribution. (d) The mode temperature of the resonator as a function of incident power for a coupled optical cavity as shown in Figure 2b. The quantum ground state temperature is also shown as a dashed strait line.

We have modelled a mg spindle shaped silicon torsional resonator which could be etched from a silicon wafer. It acts as a rigid body resonator and achieves a very good overlap to a $TEM_{01}$ optical mode as shown in Figure 3. The contours in Figure 3a show the strain amplitude, which is concentrated in the spindle ends, and minimum at the centre where the $\omega_l$ mode interacts. The strain energy ratio $\Delta E/E$ is calculated to be $6.5\times10^{-4}$ with the coating area of 0.5mm×0.5mm and thickness of 5µm. Assuming $Q_{intrinsic}$~$10^8$, $Q_{coating}$~$10^4$, the resulting Q factor of the coated resonator will be $Q_m=1.4\times10^7$. The mini-resonator torsional mode displacement distribution and the $TEM_{01}$ cavity mode amplitude distribution are shown in Figure 3b and Figure 3c respectively, which show strong overlap. We estimate that the overlap factor $B_1$~1. The effective mode temperature of the mirror as a function of input power is plotted in Figure 3d. The parameters for the mini resonator and optical cavity are given in the figure caption. When the incident power reaches 135mW the acoustic mode is cooled to the quantum ground state with an effective temperature of 48 µK if the initial temperature is 4 K.

Another challenge is to observe the mode when it is cooled to the quantum ground state. Decoherence due to the interaction with the noisy environment prevents the observation of quantum effects. In order to observe quantum effects the mean number $\bar{N}$ of acoustic oscillations in the decoherence time must be greater than unity [19], requiring

$$\bar{N} = (\frac{\hbar \omega_m^2}{k_B T_{eff} \delta_{eff}}) \geq 1 \qquad (2)$$

where $\omega_m$ is the effective resonator frequency, $T_{eff}$ and $\delta_{eff}$ are the effective temperature and the damping rate of the acoustic mode. For the system described here, the ground state temperature corresponds to $\bar{N} = 133$.

Besides opening the way to exploration of fundamental quantum mechanics in macroscopic systems, the OAPA enables measurement at the single quanta level of displacement, mass, force, charge or biological entities [20]. The phenomena can be applied on the cm scale using 3 mirror cavities, and on the km scale using simple Fabry-Perot cavities. By direct comparison with the OPA, the OAPA would be a source of phonon-photon entanglement [3] and could find applications in quantum information, teleportation, and quantum encryption.

This research was supported by the Australian Research Council and the Department of Education, Science and Training and by the U.S. National Science Foundation. We would like to thank Jesper Munch and Peter Veitch at Adelaide University and our collaborators Yanbei Chen, Phil Willems, David Reitze, Guido Muller, David Shoemaker and Gregg Harry for useful discussions. We thank the LIGO Scientific Collaboration International Advisory Committee of the Gingin High Optical Power Facility for their support.